\documentclass[twocolumn, letter]{jpsj3}
\usepackage{txfonts}
\usepackage{graphicx}
\usepackage{dcolumn}
\usepackage{bm}
\usepackage{color}
\usepackage[]{siunitx}
\newcommand{\correct}[1]{\textcolor{black}{#1}}

\title{Magnetic Properties of  YbBe$_{13}$ Probed by Neutron Scattering and Thermodynamic Measurements }

\author{Yusei Shimizu\thanks{E-mail address: yusei.shimizu.a5@tohoku.ac.jp}$^{1}$, 
 Yoichi Ikeda$^{2}$,   Toshiro Sakakibara$^{3}$,  Yoshiya Homma$^{1}$,   and   Dai Aoki$^{1}$
}
\inst{$^{1}$Institute for Materials Research (IMR), Tohoku University, Oarai, Ibaraki, 311-1313, Japan  \\
$^{2}$Institute for Materials Research (IMR), Tohoku University, Sendai, Miyagi, 980-8577, Japan  \\
$^{3}$Institute for Solid State Physics (ISSP), University of Tokyo, Kashiwa, Chiba 277-8581, Japan \\
}
\abst{
We examined the magnetic properties of YbBe$_{13}$, which exhibits an antiferromagnetic order below $T_{\rm N} \sim$ 1.2 K. 
Unlike other $M$Be$_{13}$ compounds   ($M$ = rare earth/actinide elements),  
 based on  elastic neutron scattering,  we observed  
 an incommensurate magnetic propagation vector  \mbox{\boldmath $\tau$} = (0,0,$\tau_z$), where $\tau_z$ = 0.5385 
   is  in the reciprocal lattice unit.   Additionally, we  constructed  a  precise magnetic  phase diagram    for YbBe$_{13}$.
 We observed  non-trivial magnetic anomalies in YbBe$_{13}$, which cannot be understood based on  a simple helical order.  
 Our results for YbBe$_{13}$ provide an opportunity to reconsider the   electron state of UBe$_{13}$ and  present an important step toward a comprehensive understanding of magnetic correlations in $M$Be$_{13}$ series.
 \color{black}
}

\begin{document}
\maketitle



The $M$Be$_{13}$ ($M$ = U, Th, Np, and rare-earth elements)  series with cubic NaZn$_{13}$ structure (space group: $\#$226, $O_{h}^{6}$)
   exhibits    various  physical properties and has been   extensively  investigated \cite{Bucher_PRB_1975, Ott_PRL_1983}.  \color{black} 
   Among them, UBe$_{13}$ is of particular interest as it is a system with  delocalized $f$ electrons exhibiting  extremely strong electron correlations. 
It shows  unconventional superconductivity \cite{Ott_PRL_1983},
 magnetic anomalies   ($B^{*}$ anomaly)   in the superconducting state \cite{Ellman_PRB_1991, Kromer_PRL_1998, Shimizu_PRL_2012}, 
 and non-Fermi-liquid behavior \cite{Brison_JLTP_1986, Gegenwart_PhysicaC_2004, Shimizu_PRB_2015}.
Although  UBe$_{13}$ is a promising candidate for spin-triplet  superconductors \cite{Ott_PRL_1984, Shimizu_PRL_2019}, 
   its $B^{*}$ anomaly may be  originated from a short-range spin-density-wave (SDW) state   \cite{Kromer_PRL_1998}. 
 In addition, although a magnetic correlation of ($\frac{1}{2}$, $\frac{1}{2}$, 0) has been observed in UBe$_{13}$ \cite{Coad, Hiess_PRB_2002, Hiess_PRB_2014}, no microscopic picture reflecting  the relationship between magnetic correlations and superconductivity has been obtained. 
  \color{black}

  In order to understand the basic magnetic correlations in UBe$_{13}$ with extremely strongly-correlated 
 electron state, \color{black} 
   it is also important to investigate  the detailed magnetic structure  of   $M$Be$_{13}$ in a  weakly correlated electron regime. 
The  $M$Be$_{13}$  series  tend to exhibit helical magnetism with a propagation vector of $\mbox{\boldmath $\tau$} = (0, 0, \frac{1}{3})$ ($M$ = Np, Dy, and Ho) \cite{Hiess_PRL_1996, Vigneron_JPhysF_1981, Dervenagas_PRB_2000} and a slightly modified incommensurate ordering with $\mbox{\boldmath $\tau$} = (0,0,0.285)$ for  GdBe$_{13}$  \cite{Vigneron_JPhysF_1982}, when the $f$-electrons are localized.


 In this study, we utilize  a promising  candidate, YbBe$_{13}$, to  understand  the  fundamentals of  the magnetic ground  and electronic states  when  localized $f$ electrons  acquire  itinerancy  in the $M$Be$_{13}$ series. 
 We  address the magnetic structure of YbBe$_{13}$, which    has been  reported to exhibit an antiferromagnetic order 
       below $T_{\rm N} \sim$ 1.15-1.28 K  \cite{Heinrich_PhysLett_1979, Ramirez_PRL_1986}.
 In the paramagnetic state, the magnetic susceptibility, $\chi(T)$, obeys the  Curie-Weiss law for the Yb$^{3+}$ state, \color{black}
  where the Curie-Weiss temperature   $\varTheta $  is approximately 1 K  \cite{Heinrich_PhysLett_1979}. 
   The isomer shift evaluated from M\"{o}ssbauer spectroscopy  measurements   also revealed that Yb is in the trivalent state  
  ($J = 7/2$)     with Ruderman-Kittel-Kasuya-Yosida   interaction  
  \cite{Bonville_JPhysF_1986}, although the possibility of a mixed valent state for Yb ion was discussed in an 
  earlier paper \cite{Eynatten_ZPhysB_1983}. 
  However, x-ray absorption results reported that  Yb ion  has a trivalent state in the bulk YbBe$_{13}$  \cite{Frank_JLessCommonMetals_1985}.
In YbBe$_{13}$, the crystalline-electric-field (CEF) state of  the $4f$ electrons can be  well understood from 
 inelastic-neutron scattering \cite{Walter_JMMM_1985} and heat-capacity measurements \cite{Ramirez_PRL_1986},   which suggest CEF levels of $\Gamma_{7}$-$\Gamma_{8}$-$\Gamma_{6}$, where $\Gamma_{7}$, $\Gamma_{8}$ and $\Gamma_{6}$ are the CEF ground, first-excited, and second-excited states, respectively. This indicates  that  $f$ electrons in YbBe$_{13}$ are   localized. However,  the occurrence of the Kondo effect in YbBe$_{13}$, as suggested by the M\"{o}ssbauer effect \cite{Bonville_JPhysF_1986} and heat-capacity experiments \cite{Ramirez_PRL_1986}, implies that the $f$ electrons are itinerant, with a moderate effective mass at low temperatures ($\gamma \sim$ 30 J K$^{-2}$mol$^{-1}$).
\color{black}


 To clarify   the magnetic structure of YbBe$_{13}$ in the antiferromagnetic order, we performed  elastic   neutron-scattering,  DC magnetization, and specific-heat measurements  using single-crystalline samples.  
 Based on   elastic neutron-scattering measurements, we  propose possible magnetic structures below $T_{\rm N} \sim $1.2 K,  which     have  an incommensurate propagation vector of $ \mbox{\boldmath $\tau$}  =$(0, 0, 0.5385). 
   Additionally, we  present a   non-trivial   magnetic phase diagram for  YbBe$_{13}$.


 
 Single-crystalline YbBe$_{13}$  samples were grown \correct{via}  the  Al-flux method  using alumina crucibles in a sealed quartz \correct{tube.} 
After centrifuging the Al flux, the flux on the crystal surface was removed using NaOH solution.
   No sample dependence  was observed   for single crystals obtained from the same batch.

\color{black}

 Elastic  neutron-scattering  \color{black} profiles were measured using     a     triple-axis neutron spectrometer TOPAN installed on the 6G horizontal beam port in JRR-3. The incident and scattered neutrons were monochromatized with pyrolytic graphite (PG) (002) crystals and calibrated   \correct{to}  $\lambda$ = 246.7 pm ($\sim 13.4$ meV) using an $\alpha$-alumina standard sample. The monochromator and analyzer were vertically focused\correct{,}  and collimators ($\ang{;30;}$-$\ang{;30;}$-$\ang{;30;}$-$\ang{;60;}$)   were used before and after  \correct{they were placed.}  
  \correct{High}-order harmonics were reduced using PG filters. Small single crystals (total weight\correct{:} 42 mg) were fixed on an aluminum plate using    hydrogen-free glue (CYTOP). Each crystal was coaligned such that the horizontal scattering plane was $(hhl)$. 
  A closed-cycle $^3$He refrigerator was used to cool the samples  to 0.69 K.
The mosaic of the coaligned samples  was coarse and evaluated to \correct{be}  a few degrees.  Therefore,  we focused only on determining the  magnetic reflections of YbBe$_{13}$.

\begin{figure}[htb]
\begin{center}
\includegraphics[width=6.5cm]{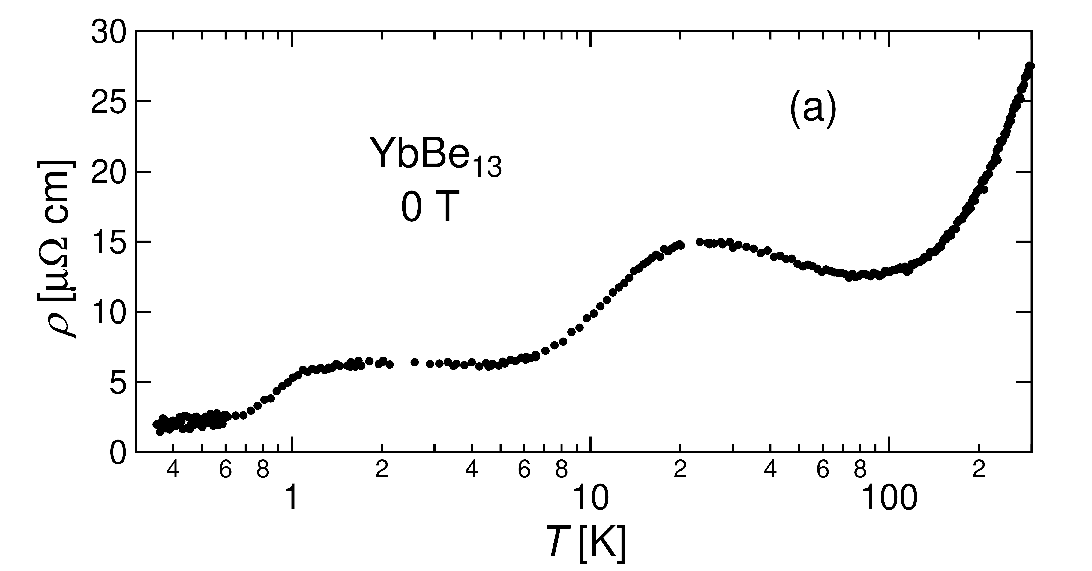}
\includegraphics[width=6.5cm]{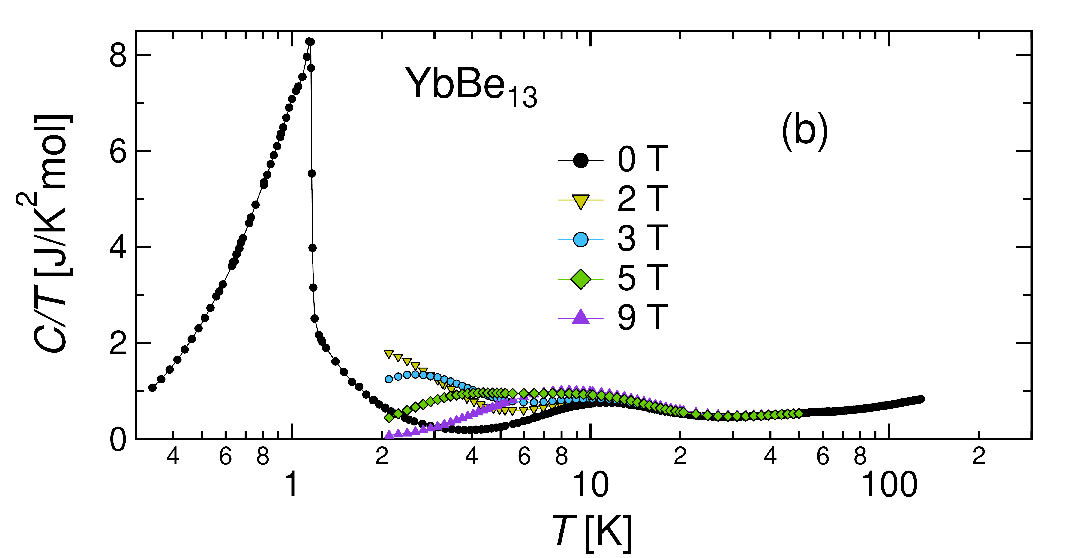}
\caption{  \color{black}  (a) Temperature dependence of resisitivty in YbBe$_{13}$ (logarithmic temperature scale),
  where the residual-resistivity ratio was 15. \color{black}
 (b)  Temperature dependence of heat capacity [$C(T)/T$] in YbBe$_{13}$ from 130 K to the lowest temperature of 0.34 K (logarithmic temperature scale).   We also plot $C(T)/T$  data below 50 K at 2, 3, 5, and 9 T along
 the  $\langle100\rangle$  axis.
 }
 \end{center}
\end{figure}

\color{black}


 Low-temperature magnetization measurements were performed \correct{using} the  Faraday method.
  We used  a high-resolution capacitively detected magnetometer \cite{Sakakibara_JJAP_1994, YShimizu_RSI_2021}
     installed in a $^{3}$He refrigerator  at low temperatures down to 0.28 K  \correct{under} magnetic fields along the cubic   
       $\langle100\rangle$  axis for a single-crystalline sample \correct{weighing} 2.68 mg.  
\correct{We examined}   the difference between the field-cooling and  zero-field-cooling processes for the antiferromagnetic order in YbBe$_{13}$ \correct{but did not observe}  significant  difference\correct{s} between them.

 Meanwhile, heat-capacity and resistivity  measurements  were performed 
  using a commercial Physical Properties Measurement Systems (PPMS Dynacool, Quantum Design)  at low temperatures down to  0.34 K.      For heat-capacity measurements, 
   \correct{we applied magnetic fields  up to 9 T} along the cubic  $\langle100\rangle$ \color{black}   axis. 
    We measured the electrical resistivity via the four-probe method  using gold wires; subsequently,  the resistivity value was normalized using a previously reported value in the literature 
    \cite{Oomi_Trans_2003}.

\begin{figure}[htb]
\begin{centering}
\includegraphics[width=8cm]{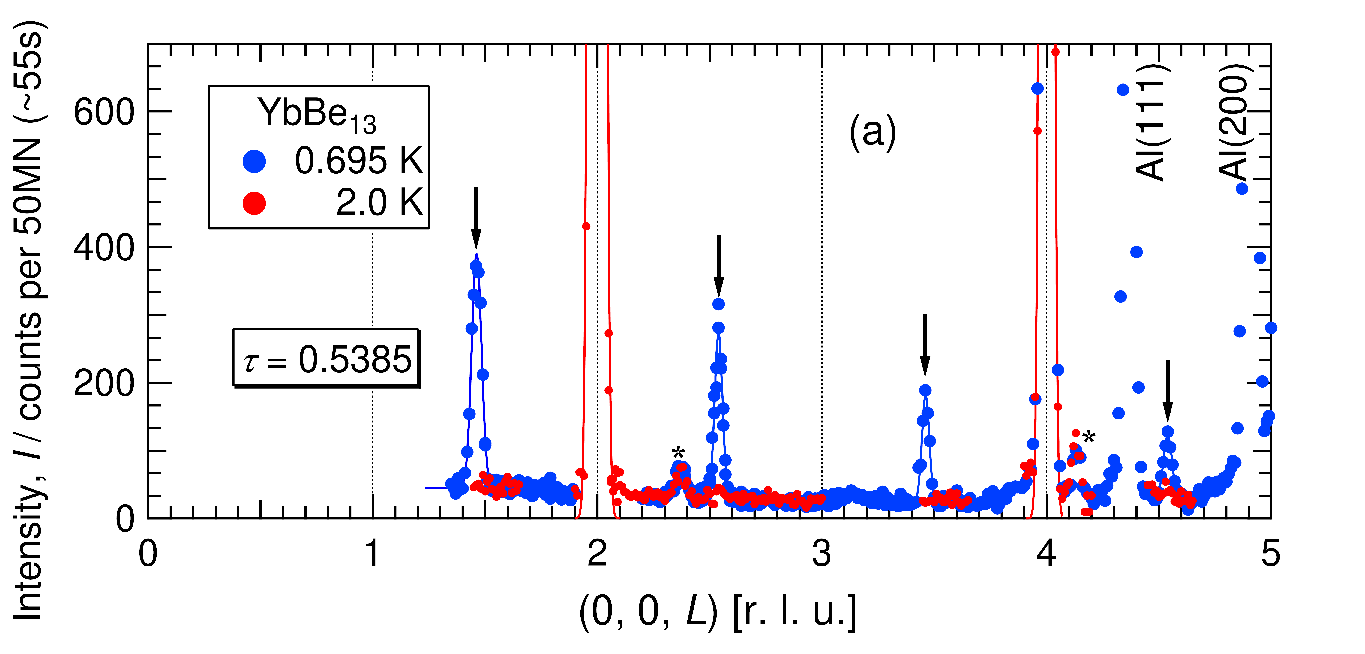}
\includegraphics[width=8cm]{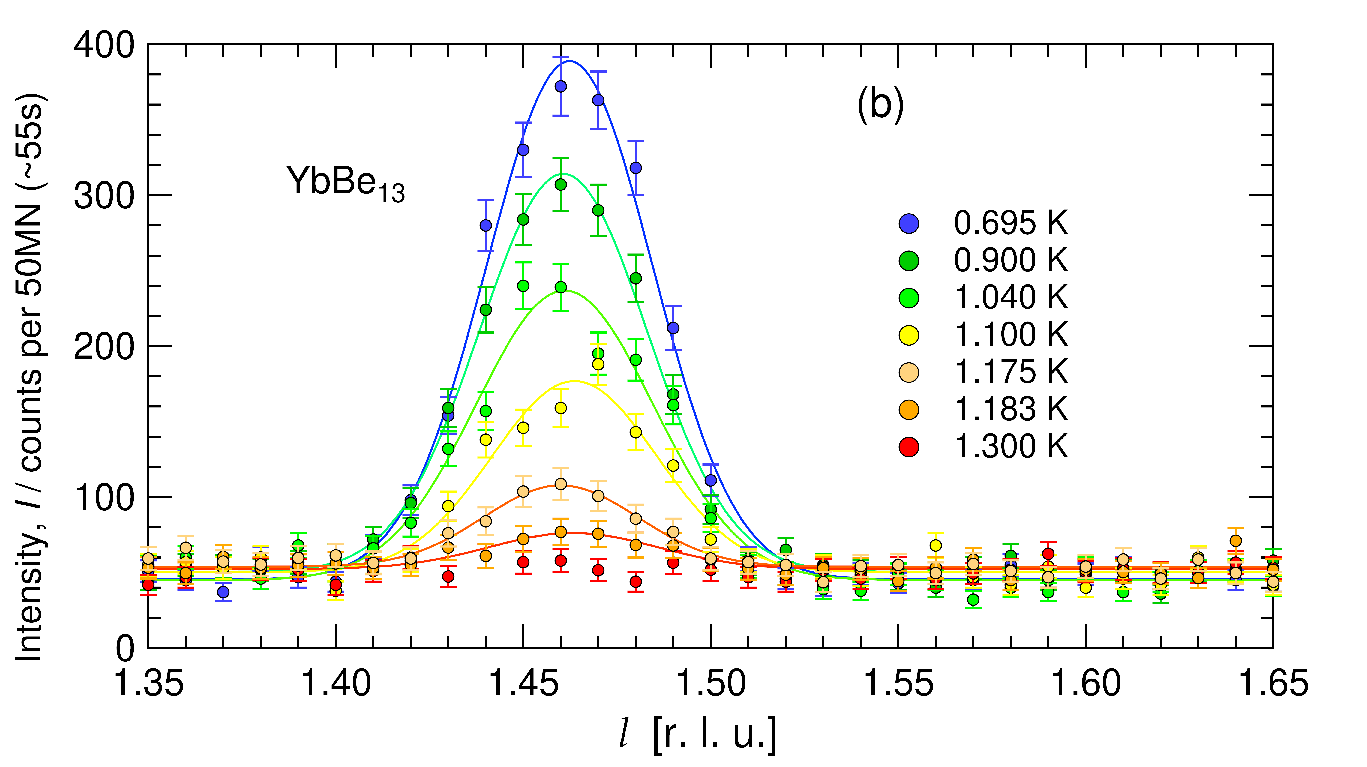}
\includegraphics[width=8cm]{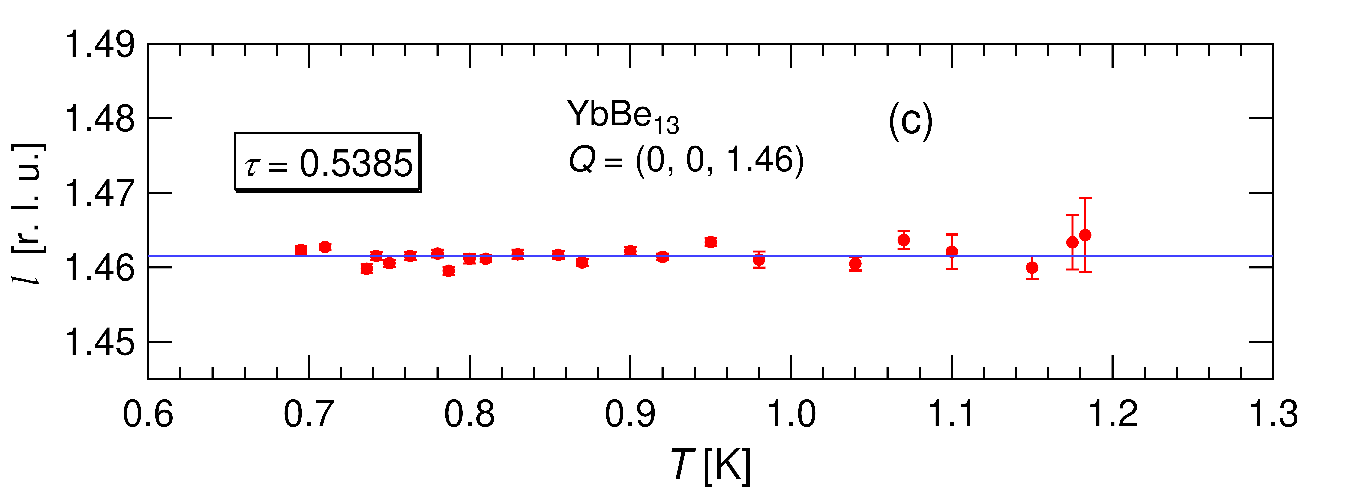}
\caption{ (a)  The line scan profile along $(00 L)$ measured at the base temperature, 0.695 K (blue circles) in the magnetic ordered state, and 2 K  in the paramagnetic state (red circles).
 (b)  Temperature dependence of  elastic neutron\correct{-}scattering intensity at  $ \mbox{\boldmath $Q$} $ = (0, 0, 1.46)
 as a function of $l$ r.l.u.
 (c) Temprature dependence of the peak centers at around  \mbox{\boldmath $Q$} = (0, 0, 1.46), where there is no change in the center of the magnetic Bragg peak within the experimental accuracy. 
 }
 \end{centering}
\end{figure}

 

\correct{First, we}  report the transport and thermophysical properties of  \correct{a}   single-crystalline YbBe$_{13}$ sample.  Figures 1(a) and 1(b)  show the temperature dependence\correct{s}  (logarithmic temperature scale) of the resistivity and heat capacity, respectively.
The resistivity decreases upon cooling from room temperature to 70 K, and there is a broad maximum \correct{at}  20-30 K  \correct{under} zero field. 
    In  the low temperature region below 20 K,   \correct{the variation of  $\rho(T)$ with respect to temperature was insignificant}  down to 2 K. At 1.18 K, a kink behavior \correct{was} clearly  
   observed in $\rho(T)$, which \correct{was}  due to the antiferromagnetic transition in YbBe$_{13}$.
 These behaviors are  consistent with  previous resistivity measurements
  \cite{Naidyuk_Physica_B_1999, Oomi_Trans_2003}.
 \correct{Meanwhile, the $\rho(T)$  data suggest } that the  $4f$ electrons in YbBe$_{13}$ are not in a  strongly correlated regime,  \correct{as $\rho(T)$ varies weakly}  and the \correct{value of}  $A$ coefficient is very small at low temperatures.

   
In the heat\correct{-}capacity data measured at zero field,   a broad local maximum  \correct{was observed } at   10 K, which \correct{is approximately}   half  the  value of  the resistivity maximum temperature. There is  no sample dependence for the observed broad maximum behavior in the resistivity and heat capacity of the   single-crystalline samples. 
\correct{Meanwhile,} 
 $C(T)/T$ exhibit\correct{ed}  a large $\lambda$-shaped anomaly at $T_{\rm N} \sim $ 1.2 K owing to the antiferromagnetic transition,  \correct{thus} suggesting  \correct{a second-order} phase transition.  Above $T_{\rm N} $, a large enhancement in  $C/T$  is observed,  presumably \correct{owing}  to  antiferromagnetic fluctuations. 
\correct{The} heat-capacity data obtained \correct{under}  magnetic fields up to 9 T \correct{were} plotted, 
 \correct{as  shown in Fig. 1(b).}
 The broad  anomaly at 10 K 
 \correct{varied significantly under}  the applied magnetic fields.  Moreover,  an additional peak   anomaly \correct{appeared}  at 3-4 K at  3 and \correct{5 T.} 
 Previous inelastic neutron\correct{-}scattering \cite{Walter_JMMM_1985} and heat\correct{-}capacity  \cite{Ramirez_PRL_1986} studies suggested that the CEF ground state is the  $\Gamma_{7}$ doublet, 
  \correct{whereas} the first and second excited states  are  the  $\Gamma_{8}$ quartet  and  $\Gamma_{6}$ doublet, respectively.  The  field-induced   anomaly observed at 3 and 5 T may originate from the Schottky anomaly with a splitting of the $\Gamma_{7}$ ground state in the magnetic field.

\begin{figure}[htb]
\begin{center}
\includegraphics[width=8.3cm]{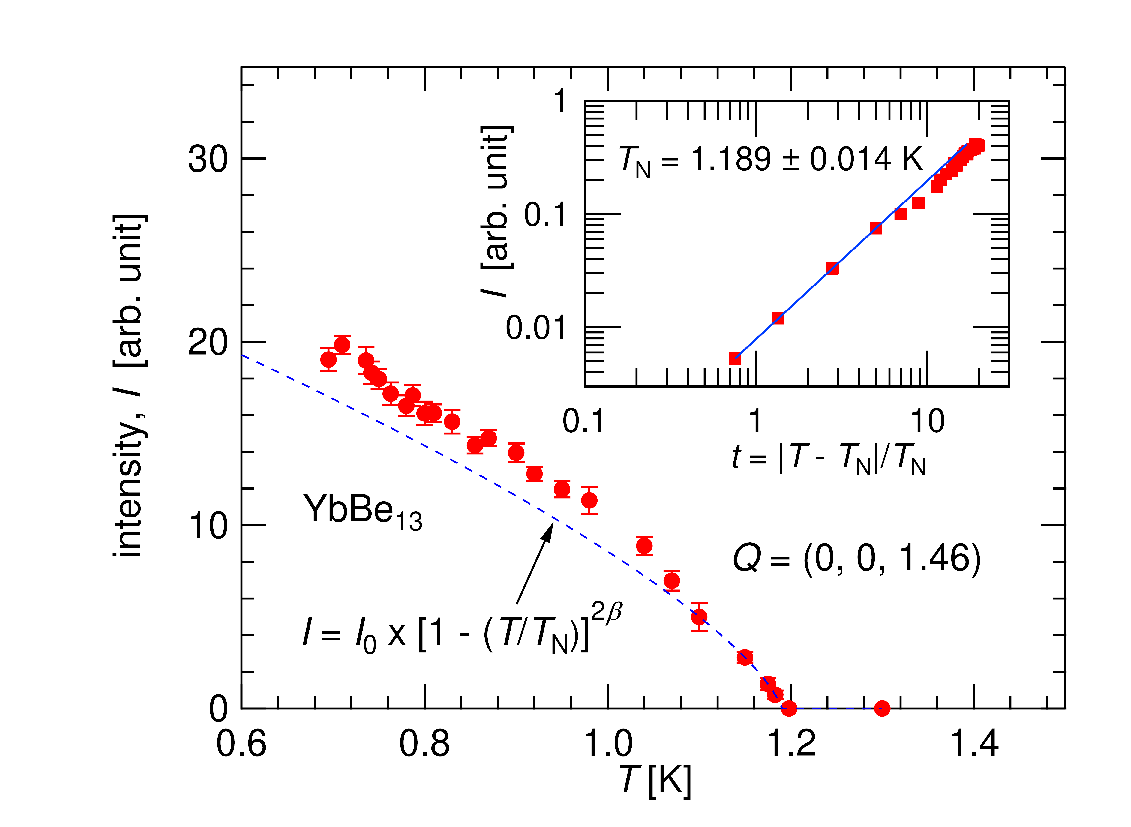}
\vspace*{10pt}
\end{center}
\caption{  Neutron\correct{-}scattering intensity  ($I$) in YbBe$_{13}$ at $ \mbox{\boldmath $Q$} $  = (0, 0, 1.46)  as a function of temperature. 
 The dashed line is  the result of a fit with  $I  = I_{0} \times [1 - (T/T_{\rm N})]^{2 \beta} $,
   where $T_{\rm N}$ = 1.189  K and $\beta$ = 0.36.  
The inset shows the log-log plot of $I$ versus the  reduced temperature $t \equiv | T - T_{\rm N} |/T_{\rm N}$.
 }
\end{figure}

Next, we report  the results of  \correct{elastic} neutron\correct{-}scattering measurements for YbBe$_{13}$.
 Several line scans  were  \correct{performed}    to determine the magnetic reflections  below $T_{\rm N}$.
 A typical line\correct{-}scan profile is shown in Fig. 2(a). The blue and red points were measured at  0.695 K (below $T_{\rm N}$)  and 2 K (above $T_{\rm N}$), respectively.   The intense peaks around $l$ = 2 and 4 r. l. u. are the fundamental nuclear reflections from YbBe$_{13}$,  \correct{whereas} those around $l \sim$ 4.3 and 4.8 r. l. u. are background Bragg reflections from the aluminum cell.
  Additionally,  the temperature-independent small peaks around $l$  $\sim$ 2.4 and 4.1 r. l. u. are spurious scatterings from the refrigerator. The  \correct{absence} of clear Bragg reflections at \mbox{\boldmath $Q$} $= (0, 0, 2n+1)$, where $n$ is  an integer, is consistent with  \correct{the} face-centered cubic structure of YbBe$_{13}$  
    (space group $\#226$, $Fm\bar{3}c$).  
   At the base temperature,  clear satellite peaks are observed around the  (002) and (004) \correct{planes},
    as  indicated by  the downward black arrows in Fig. 2(a). 
  These satellites can be expressed as  \mbox{\boldmath $Q$}  $=$ \mbox{\boldmath $G$} $\pm$ \mbox{\boldmath $\tau$}, where  \mbox{\boldmath $G$}  is the reciprocal lattice vector of YbBe$_{13 }$ and $\mbox{\boldmath $\tau$} = (0, 0, \tau_{z})$ is the magnetic propagation wave vector. To evaluate $\tau_z$, the observed peaks were fitted with a Gaussian function, and  $\tau_z$ was evaluated to be 0.5385 $\pm$ 0.0003 r.l.u. at the base temperature. This incommensurate propagation vector in YbBe$_{13}$ \correct{differs} completely  from that of    \correct{the} commensurate magnetism   \correct{in} $M$Be$_{13}$ ($M$ = Dy, Ho\correct{,}  and Np) \cite{Hiess_PRL_1996, Vigneron_JPhysF_1981, Dervenagas_PRB_2000}.

 To determine whether the observed peak \correct{is} due to a magnetic transition, we examined  the temperature evolution of the peak intensity \correct{near} \mbox{\boldmath $Q$}  = (0, 0, 1.46). 
As shown  in Fig. 2(b), the observed peak \correct{became smaller with  increasing} temperature and fully \correct{disappeared} at $T$  = 1.3 K (above $T_{\rm N}$). Furthermore, the observed intensity \correct{shown} in Fig. 2(a) \correct{decreased as}
    the  scattering vector \correct{increased}. The \mbox{\boldmath $Q$}  dependence can be interpreted as  the  \correct{effect} of the magnetic form factor of  \correct{the} Yb ions. These results indicate that the observed satellites are most likely due to the magnetic transition below $T \sim$  1.2 K, which is consistent with the results of $\rho (T)$ and $C(T)$.

    Figure 3 shows the temperature dependence of the neutron\correct{-}scattering intensity at \mbox{\boldmath $Q$}  =   \correct{(0, 0,} 1.46)  for  YbBe$_{13}$. The  neutron intensity ($I$) \correct{clearly developed} below $T \sim $ 1.2 K,     \correct{which is consistent} with the heat\correct{-}capacity results \correct{ [Fig. 1(b)].}   
    The dashed line \correct{presents}  the \correct{fitting} between  1.1  and 1.2 K  with an empirical formula, $I  = I_{0} \times [1 - (T/T_{\rm N})]^{2 \beta} $, where $T_{\rm N}$ = 1.189 $\pm$ 0.014 K and $\beta$ = 0.36 $\pm$ 0.25.     The inset shows \correct{$I$ vs.}  the  reduced temperature $t \equiv | T - T_{\rm N} |/T_{\rm N}$ in the logarithmic scale.
 The neutron\correct{-}scattering intensity is generally proportional to the volume of the specimen ($V$) and the square of the  ordered magnetic moment   ($m^{2}$), i.e.,   $I  \propto V m^{2}$.  In the case of  mean-field approximation,  $\beta$ is 0.5, \correct{whereas it}  is 0.35 and 0.36 for the $XY$  \correct{and three-dimensional} Heisenberg model\correct{s}, respectively   \cite{Chaikin_Lubensky}.    These values of  $\beta$   are not significantly different from the experimentally determined value.

\begin{centering}
\begin{figure}[t]
\includegraphics[width=8.1cm]{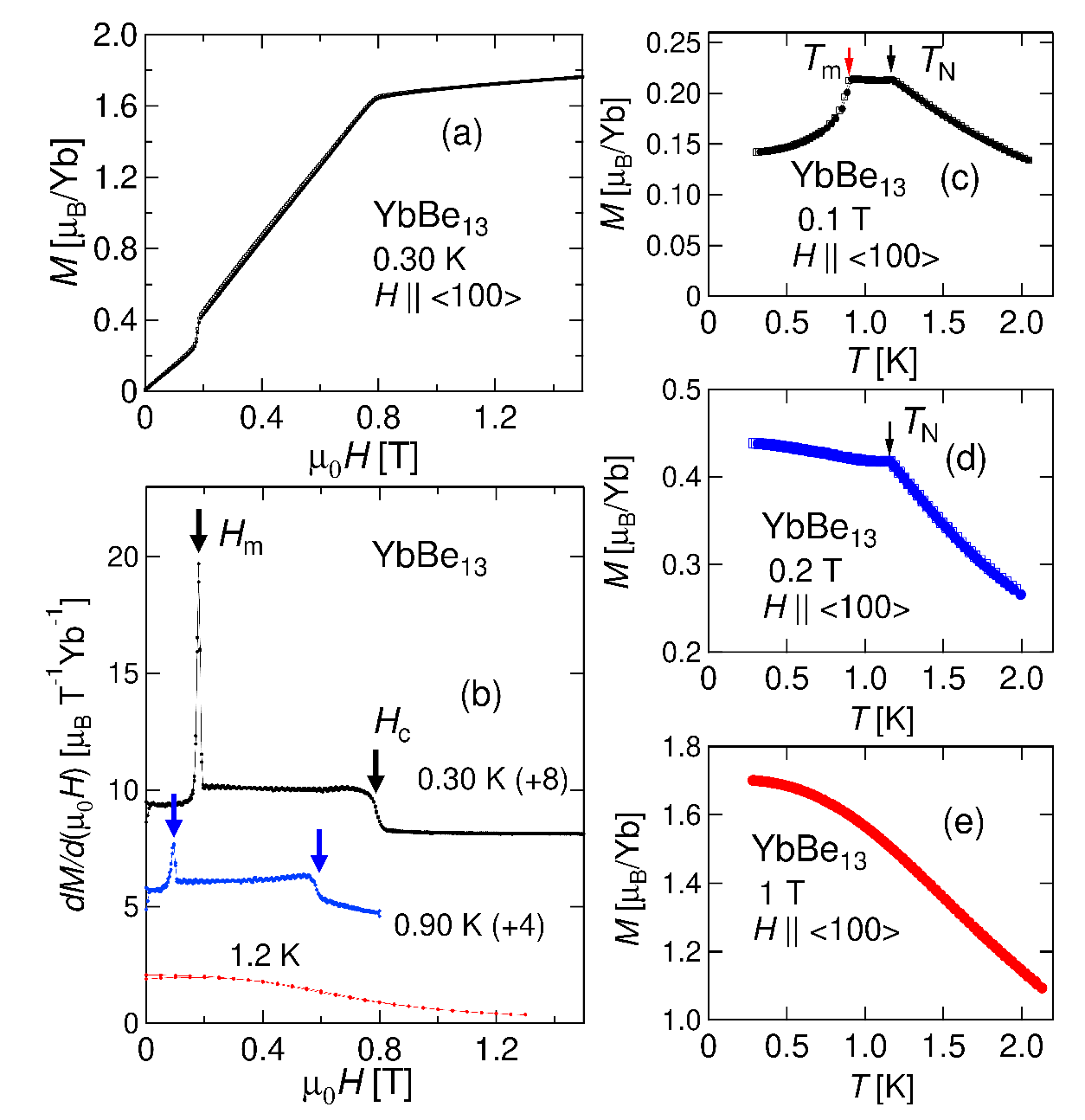}
\vspace*{20pt}
\caption{ \color{black}  (a) Magnetization process of YbBe$_{13}$, measured  in magnetic fields applied  along the cubic $\langle100\rangle$  axis.   (b) Differential magnetization of YbBe$_{13}$, $dM/dH$, measured at 0.30, 0.90, and 1.2 K.  
 Here, the data measured at 0.30 and 0.90 K are shifted vertically for the clarity, where the values are denoted in parentheses.  
  Temperature dependence of magnetization, measured at (c) 0.1, (d) 0.2, and (e) 1 T,   for    $H$ $ || $ $\langle100\rangle$.
 }
 \label{f2}
\end{figure}
\end{centering}

   \correct{The p}ossible magnetic structures can be listed \correct{via an}  irreducible representation analysis 
   \correct{based on} the evaluated propagation vector  \cite{Bertaut_ActaCryst_1968, Izyumov_JMMM_1979}. 
     For YbBe$_{13}$, only three types  of magnetic structures can be realized just  below $T_{\rm N}$ for $ \mbox{\boldmath $\tau$}  = $(0, 0, 0.5385). 
     Two of these \correct{exhibit}  longitudinal sinusoidal modulations along the 
        $[001]$  \color{black}axis.
      The third \correct{type is} helicoidal modulation, in which  the magnetic moments are perpendicular to the propagation vector ($[001]$  axis). 
We will discuss the magnetic\correct{ally} ordered state  in YbBe$_{13}$ later\correct{, in addition to} the low-$T$ magnetization data.

    Figure 4(a) shows the magnetization process of YbBe$_{13}$ in magnetic fields along the cubic 
     $\langle100\rangle$   axis, measured at 0.30 K.   A clear step-like behavior is observed in $M(H)$  at $\mu_{0} H_{\rm m} =$ 0.18 T,  and a kink appears at $\mu_{0} H_{\rm c} =$ 0.8 T. 
    Above 0.18 T, the magnetization curve is almost linear \correct{with}  $H$
   below \color{black} $\mu_{0} H_{\rm c} = $ 0.8 T.  An  anomaly is observed  at $H_{\rm c}  $ in both the increasing and decreasing field processes, \correct{thus}  implying  a phase transition.
  In \correct{this} study, we performed magnetization measurements \correct{under fields} up to  6 T\correct{;   
   however,} no anomalies were observed  above  $H_{\rm c}$. 
     Figure 4(b) shows the differential magnetization [$dM(H)/dH$] \correct{curves} obtained at  0.30, 0.60, and 1.2 K.  \correct{The} sharp cusp in $dM(H)/dH$  at $H_{\rm m}$ indicates a first-order transition. 
      \correct{Meanwhile, the} step-like behavior  in $dM(H)/dH$ at $\mu_{0} H_{\rm c} =$ 0.8 T (0.6 T) for 0.30 K (0.90 K) suggests a second-order phase transition at $H_{\rm c}$.

\begin{centering}
\begin{figure}[t]
\includegraphics[width=4.2cm]{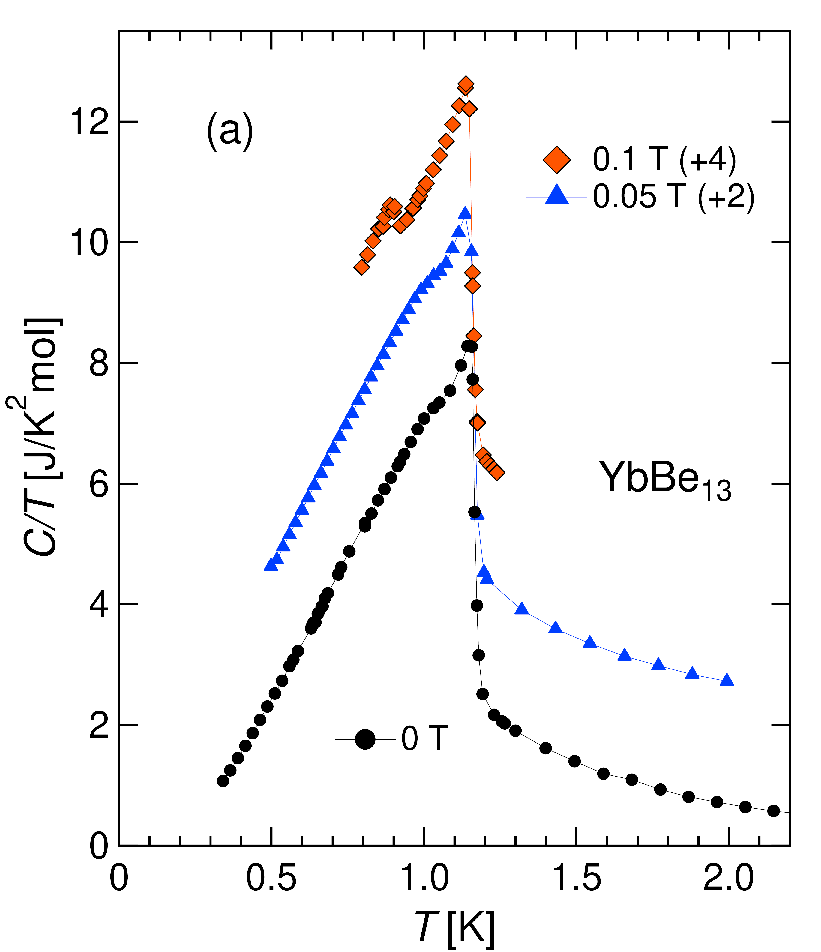}
\includegraphics[width=4.2cm]{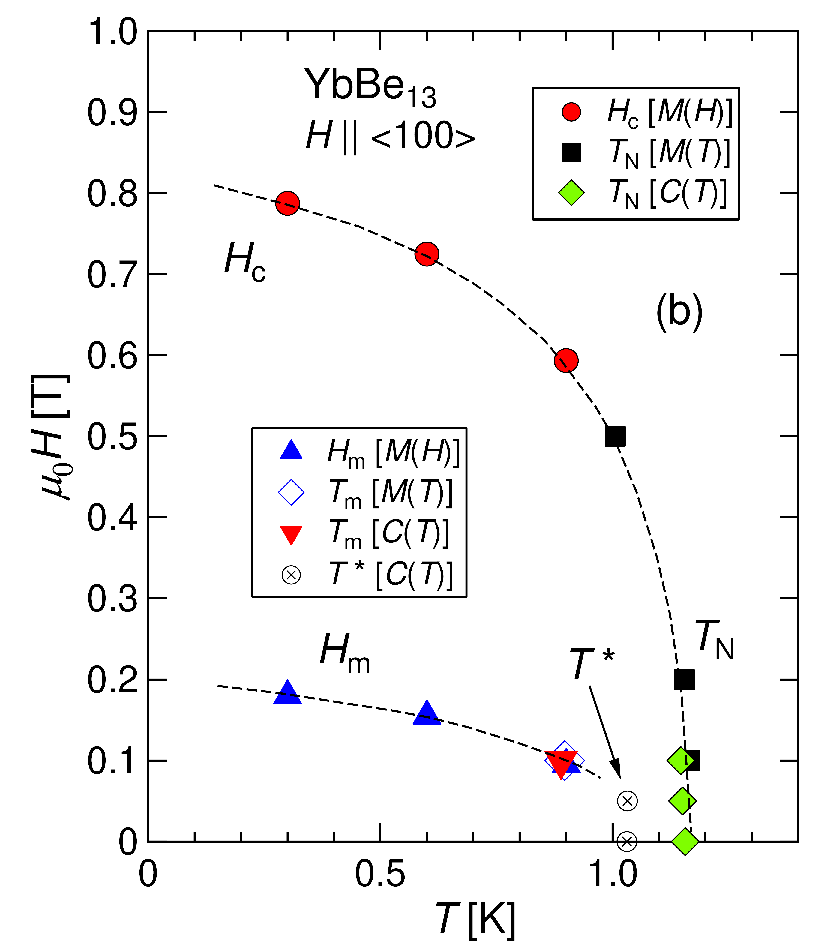}
\caption{  (a) The temperature dependence of heat capacity in single crystalline YbBe$_{13}$, measured at zero and magnetic fields, 0.05 and  0.1 T applied along the cubic $\langle100\rangle$ axis. Here, the data measured in magnetic fields are shifted vertically for the clarity, where the values are denoted in parentheses.  (b) The obtained $H$-$T$ phase diagram in YbBe$_{13}$ for $H$ $||$   $\langle100\rangle$.  The details are explained in the text. 
 }
 \label{f3}
\end{figure}
\end{centering}

\correct{Next, we}  examined  the temperature dependence of \correct{the} magnetization [$M(T)$]. Figures 4(c), 4(d),  and  4(e) show  $M(T)$,  measured at 0.1, 0.2, and 1 T, respectively. 
 At 0.1 T,   a distinct  anomaly below 1 K is  apparent,  which is denoted as   $T_{\rm m}$ [Fig. 4(c)], \correct{in both the}  increasing and decreasing processes.  This  behavior is  \correct{consistent} with the magnetic\correct{-}field dependence of  $M(H)$.  At 0.2 T,    such behavior \correct{was not observed  below $T_{\rm N}$.} 
    At 1 T,  \correct{which is} higher than the critical field ($H_{\rm c}$) of the antiferromagnetic order in YbBe$_{13}$, no anomalies \correct{were} observed in $M(T)$.

Figure 5(a) shows the temperature dependence of the heat capacity  measured at zero and 
 \correct{under} magnetic fields of 0.05 and 0.1 T. 
   At zero field, a large peak \correct{was} observed at $T_{\rm N} \sim $1.2 K owing to the antiferromagnetic transition. The observed behavior is similar to that  reported \correct{previously for}   heat capacity \cite{Ramirez_PRL_1986}.
     However,  in previous studies, \correct{shoulder-like  anomalies} at $\sim$0.9-1 K \correct{were} not very \correct{evident}.
  \correct{In this study, we} observed a  shoulder-like anomaly ($T^{*} \sim$0.9-1 K)  under a very  low field below 0.05 T, which is \correct{likely} a crossover  anomaly in the antiferromagnetic phase.  
 \correct{Notably,} the  relationship  \correct{among}  $H_{\rm m}$, $T^{*}$, and $T_{\rm N}$ in the $H$-$T$ phase diagram of YbBe$_{13}$ \correct{is worth investigating in future studies}.

Figure 5(b) shows the $H$-$T$ phase diagram of YbBe$_{13}$ for $H$ $||$ 
 $\langle100\rangle$\color{black}.
  The open circles, squares, and diamonds represent the critical fields of antiferromagnetic order in YbBe$_{13}$
   \correct{o}btained from  the magnetization process [$M(H)$], \correct{the} temperature dependence of magnetization [$M(T)$],  and the temperature 
   dependence of \correct{the} heat capacity, respectively. 
   \correct{Additionally, we} plotted  the \correct{corresponding anomalies} inside the antiferromagnetic phase, 
   \correct{as denoted} by $M(H)$ (triangles), $M(T)$ (diamonds), and $C(T)/T$ (circles with crosses).
        In  the low-field region,  anomalies \correct{were} evident in  the field and temperature dependences of the magnetization, which are denoted as $H_{\rm m}$ and  $T_{\rm m}$, respectively. 
      This low-field anomaly can be similarly explained  
     based on    $C(T)/T$  data at 0.1 T [Fig. 5(b)].\color{black}

Having established that the magnetic order in YbBe$_{13}$ has a propagation vector of  $\mbox{\boldmath $\tau$}  = $ (0, 0, 0.5385),
  \correct{we}  now  discuss the behavior of the  magnetization curve \correct{for}  YbBe$_{13}$. 
    In a cubic system, \correct{three}  equivalent domains with  propagation vectors of   ($\tau_{x}$, 0, 0),  (0, $\tau_{y}$, 0), and (0, 0, $\tau_{z}$) \correct{should exist} at zero  field. 
 Such a multi-domain state is destabilized in a finite magnetic field because of the non-equivalent Zeeman effect between the  parallel and orthogonal domains \cite{KitanoNagamiya_PTP_1964}.
       In GdBe$_{13}$, a multi-domain state \correct{was} discussed in  the low-field region  of its helical order \cite{Hidaka_PRB_2020}. 
   \correct{By} contrast, a change in the domain structure cannot  explain  the $M(H)$ data of YbBe$_{13}$ 
  because the anomaly at $H_{\rm m}$ is  evident in the field-decreasing  process of $M(H)$ [\correct{Fig.} 4(a)],  unlike in the case of GdBe$_{13}$.

   Hence,  a  plausible explanation for the observed  anomalies at $H_{\rm m}$ and  $T_{\rm m}$  is 
     a   change in the magnetic structure  \correct{of} YbBe$_{13}$.
 In the case of  helical order  with $\mbox{\boldmath $\tau$}  = $ (0, 0, 0.5385) with a helical (001) planes,  
  the helical state  becomes a conical state with \correct{a} linear $M(H)$ curve, unlike  the 
   \correct{case for the }experimental $M(H)$
   data in YbBe$_{13}$.
   Although the linear   $M(H)$ curve in YbBe$_{13}$ above $H_{\rm m}$ is  \correct{ similar}   
 to  that of helical magnetic  systems  \cite{Hidaka_PRB_2020, Komatsubara_JPSJ_1970, Sakakibara_JPSJ_2021, Bauer_PRM_2022},
  the low-field ordered state below the first-order transition  at $H_{\rm m}$ cannot be  explained by a simple helical order.
The determination of the  precise magnetic structure of YbBe$_{13}$ is  currently in progress.  
\color{black}


\correct{Under 0.05 T  and} zero field, a shoulder-like anomaly \correct{was} observed  at $T^{*} \sim $ 1 K in 
  $C(T)/T$ curves   [Figs. 5(a) and 5(b)]. \color{black}
   The entropy change for this anomaly is very small. 
 Therefore, a \correct{significant} change in the magnetic structure may be excluded at $T^{*}$.
   In connection with this weak anomaly,  no clear sign is observed  in the resistivity within the experimental accuracy. 
Although the temperature evolution of the neutron\correct{-}scattering intensity \correct{was}  reproduced 
\correct{approximately using}  the empirical equation,      
  the neutron\correct{-}scattering intensity at   $ \mbox{\boldmath $Q$}   =$ (0, 0, 1.46)  \correct{appeared}  to be slightly enhanced   below   
   1 K  (Fig. 3),  at  which the neutron intensity $I$ \correct{deviated} from the critical behavior in \correct{the}  
    log-log plot (inset of Fig. 3). 
\correct{Hence, } the modification of the magnetic structure 
  \correct{due to} the slight change in the coefficients of the linear combination for the basis vectors
   \correct{might cause the weak anomaly at $T^{*}$   in YbBe$_{13}$}.
  \correct{In fact,}   an enigmatic helical magnetic structure with sine modulation  in NpBe$_{13}$ \correct{suggests}  a complex magnetic correlation in $M$Be$_{13}$ \cite{Hiess_PRL_1996}.
  In addition, 
    based on the emergence of third-order harmonics at low temperatures,
     \correct{ a  rearrangement of   the magnetic structure in HoBe$_{13}$ has been reported}    \cite{Dervenagas_PRB_2000}.
    \color{black}
  Further studies are necessary to understand the observed anomalies in the incommensurate order in YbBe$_{13}$.


Finally, we discuss the reason for  the \correct{incommensurate} propagation  (0, 0, 0.5385)  vector 
 \correct{appearing} in YbBe$_{13}$,
  which \correct{differs significantly}  from the commensurate vector  (0, 0, $\frac{1}{3}$)  observed  in other    $M$Be$_{13}$ compounds  ($M$ = Np, Dy, and Ho) \cite{Hiess_PRL_1996, Vigneron_JPhysF_1981, Dervenagas_PRB_2000}.    
  An incommensurate magnetic order is a typical  characteristic of the  
  SDW  state and   is formed on the Fermi surface.  
    Therefore,  the Fermi surface of YbBe$_{13}$ must be clarified and the manner by which an incommensurate magnetic correlation  emerged must be elucidated.  \color{black} 
  In YbBe$_{13}$,  the 
     characteristic resistivity $\rho(T)$  above $T_{\rm max} \sim$30 K  indicates the Kondo effect [Fig. 1(a)], 
 which is consistent with   previous inelastic-neutron scattering \cite{Walter_JMMM_1985}, 
   the M\"{o}ssbauer effect \cite{Bonville_JPhysF_1986},  \color{black} 
  and   point-contact spectroscopy  experiments \cite{Nowack_PRB_1997}. 
  However, 
 the  evaluated  Kondo temperature  $T_{\rm K} \sim $ 1 K    \cite{Walter_JMMM_1985, Besnus_JMMM_1992} is much smaller than $T_{\rm max}$, thus implying that the conduction electrons are  affected  by  the first-excited CEF state in YbBe$_{13}$.
  Such a small $T_{\rm K} $ and  the resistivity anomaly at approximately 30 K, which may be related to   the CEF excited state, have  been similarly observed  in UBe$_{13}$
   \cite{Ott_PRL_1983}.
   Our results for YbBe$_{13}$ provide an opportunity to fundamentally reconsider the enigmatic $5f$ electron state  and $B^*$ anomaly in UBe$_{13}$.



In summary, we examined the  magnetic properties of YbBe$_{13}$ \correct{via} elastic neutron\correct{-}scattering, DC magnetization
  and heat\correct{-}capacity measurements using single-crystalline samples. 
  \correct{Based on} elastic neutron\correct{-}scattering measurements,   an incommensurate propagation  
   \correct{vector}   $\bm{\tau} = $(0, 0, 0.5385)  was  unexpectedly observed  below   $T_{\rm N} $,
    unlike other $M$Be$_{13}$ compounds.  
 Furthermore, we obtained   a non-trivial magnetic phase diagram in YbBe$_{13}$. 
  Our  findings  for YbBe$_{13}$ provide new insights  into magnetic  correlations in  the $M$Be$_{13}$  series   when $f$ electrons acquire an itinerant nature.
 \color{black}

\color{black}

 We  would like to thank  Manabu Okawara for technical support in JRR-3 in Tokai.
 The neutron scattering experiments were performed under the Joint-Use Research Program for Neutron Scattering, Institute for Solid State Physics (ISSP), University of Tokyo, at the Japan Research Reactor JRR-3 (6G IRT program 23402) and a part of the GIMRT research program in IMR (202212-CNKXX-0402). We gratefully acknowledge the support on the neutron experiments from the Center of Neutron Science for Advanced Materials, IMR, Tohoku University.
  This work was partly supported by Graints-in-Aid KAKENHI (No. 20K03851, 23K03314, 	22KK0224,	23H04870) from the Ministry of Education, Culture, Sports, Science and Technology (MEXT) of Japan. 
  The magnetization measurements were conducted  under  the Joint-Use Research Program in the Institute for Solid State Physics, the University of Tokyo.


\bibliography{apssamp}

\end{document}